\documentclass[11pt]{article} 
\usepackage{moriond,epsfig}

\def\be{\begin{equation}} 
\def\ee{\end{equation}} 
\def\bea{\begin{eqnarray}} 
\def\eea{\end{eqnarray}} 
 
\begin{document} 
\vspace*{4cm} 
\title{A JOINT FIT TO THE COSMOLOGICAL MODEL AND THE $T-M$ AND $L_x-T$ 
       RELATIONS USING CLUSTER DATA} 
 
\author{J.M. DIEGO$^{1,2}$, E. MARTINEZ-GONZALEZ$^1$, J.L. SANZ$^1$, J. 
 SILK$^2$, L. CAYON$^1$, N. BENITEZ$^3$} 
 
\address{
$^1$ IFCA, Avda. Los Castros s/n. 39005 Santander, Spain. \\
$^2$ NAPL, Astrophysics, Keble Road, Oxford OX1 3RH, UK.\\
$^3$ Dept. Physics \& Astronomy. The Johns Hopkins Univ. 3400 North 
Charles Street, Baltimore, MD 21218-2686. US  } 
 
\maketitle\abstracts{In this work we show how well can the cosmological 
parameters be constrained using galaxy cluster data. 
We also show how in the process of fitting the model 
which attempts to describe the data it is possible 
to get some information about the cluster scaling relations 
$T-M$ and $L_x-T$. Among other conclusions we found that 
only low density universes ($\Omega \approx 0.3$ with or 
without $\Lambda$) are compatible with recent data sets. 
These constraints will be improved with future SZ data.} 
 
\section{Introduction} 

Galaxy clusters have been widely used as cosmological probes. 
The strong dependence of the cluster mass function with the 
cosmological model and more particularly of its evolution 
with redshift makes the study of the cluster population a powerful 
tool to discriminate among different structure formation scenarios.
Following this fact, many works have tried to constrain the 
cosmological parameters by, given a cosmological model,  
comparing the theoretical predictions of the cluster mass 
function (Press-Schechter formalism (PS) (Press \& Schechter 1974) 
or N-body simulations) with 
cluster data obtained mainly from the X-ray band.\\
Those theoretical predictions typically provide the cluster 
abundance as a function of mass and redshift. Redshift estimates 
can be obtained from optical observations of the cluster and even 
from X-ray spectroscopic considerations if the  data is good enough. 
But the situation is different with mass estimates. Galaxy cluster 
masses are uncertain for most of the clusters and error bars are 
typically of the order of $20 \%$ or higher. For this reason, 
it is preferable to use another better determined function 
different from the mass function to trace the population of galaxy 
clusters. This can be done for instance 
by using the luminosity function of X-ray clusters or the temperature 
function. These functions can be easily connected with the theoretical 
(and model dependent) mass function  (e.g. PS formalism) by 
defining some cluster scaling relations, $T-M$ and $L_x-T$, 
\begin{equation}
 \frac{dN(T,z)}{dV(z)dT} =\frac{dN(M,z)}{dV(z)dM}\frac{dM}{dT},  
\label{dNT}
\end{equation}
where $dN(T,z)/dV(z)dT$ is the temperature function and  
$dN(M,z)/dV(z)dM$ is the mass function given for instance by PS.
Therefore, if there is an estimate of the cluster temperature 
function (or the X-ray cluster luminosity function), it would be 
possible to compare such an estimate with the predictions given by 
any model.\\
In this work we will follow this idea but with two important differences 
with respect to other previous similar works.\\
As a first important point we will consider different data sets 
in our fit and not only one as usual. This is important since the best 
fitting model will be compatible with different data sets and will reduce 
the degeneracy in the parameters found when only one data set is used.\\
Our main second difference is that in building the theoretical 
temperature, X-ray luminosity and flux functions we will 
consider that the $T-M$ and $L_x-T$ relations are not fixed relations 
but they will be part of our fitting model. This point will prevent us 
of doing wrong assumptions about these relations. 
%%%%%%%%%%%%%%%%%%
\section{Results}%
\begin{figure}
\begin{center} 
\psfig{figure=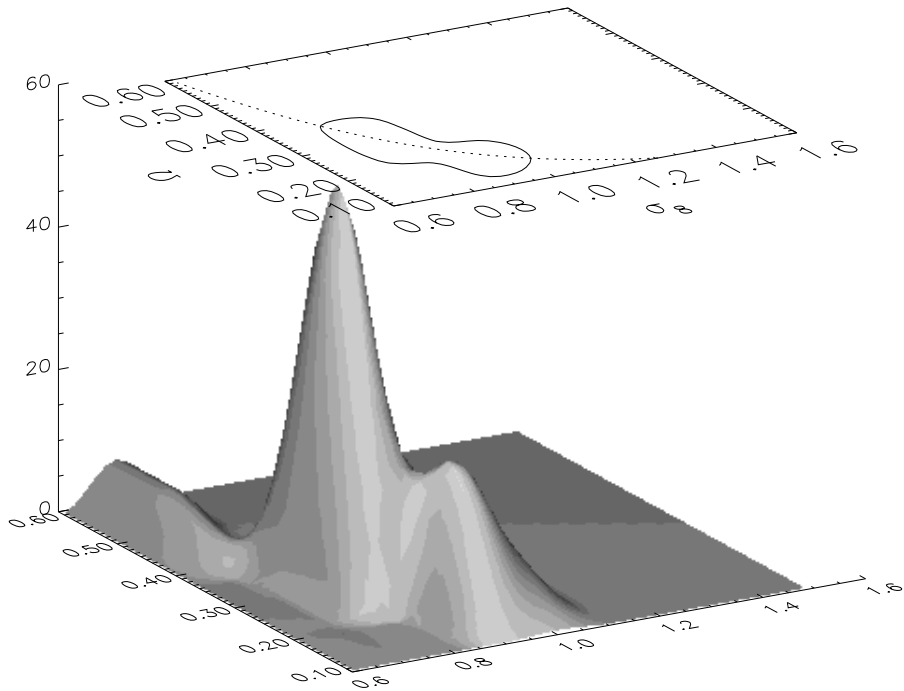,height=4.0in} 
\caption{\label{fig_Sigma8_Omega}
         Marginalized probability in $\sigma_8-\Omega_m$ for the flat  
         $\Lambda$CDM case.} 
\end{center}
\end{figure}

\begin{figure}
\begin{center} 
\psfig{figure=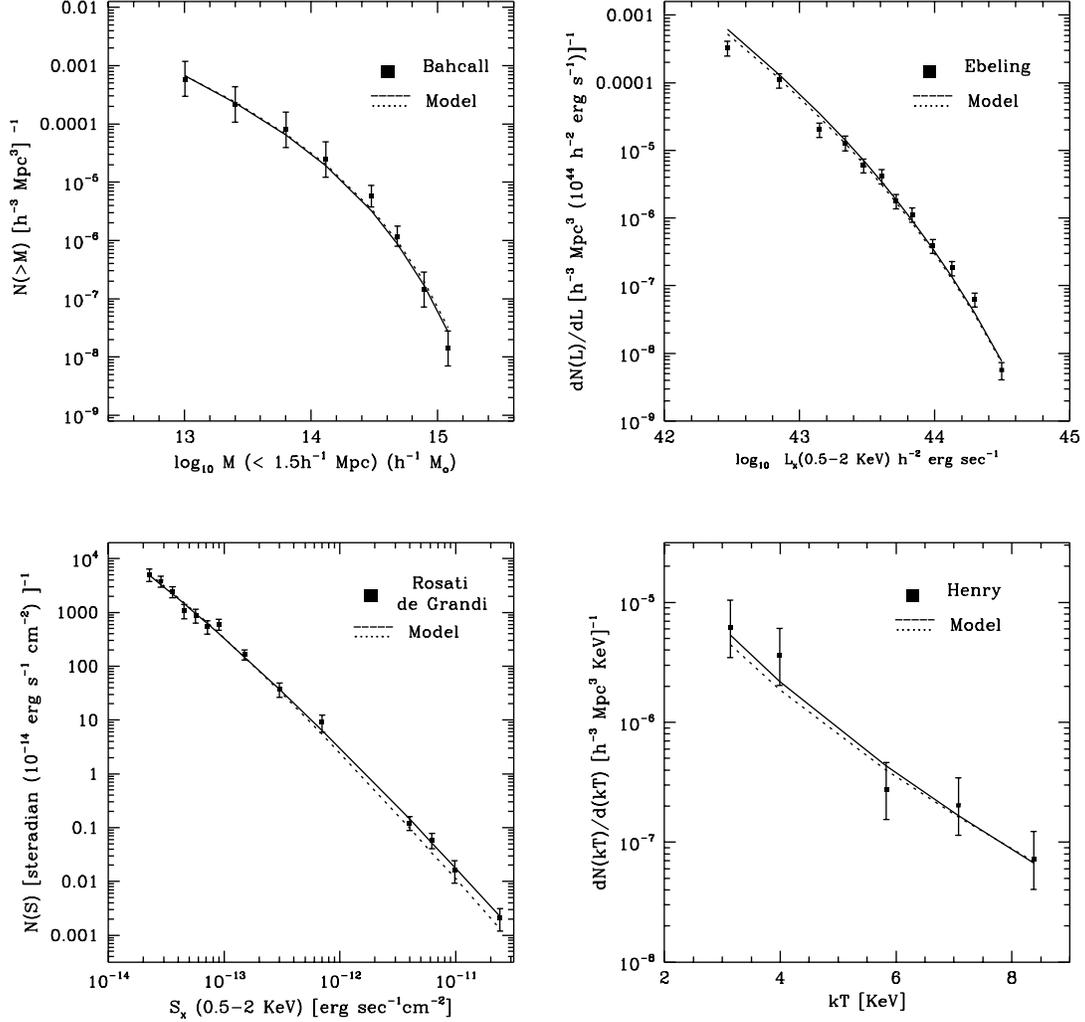,height=6.0in} 
\caption{\label{fig_best_4curves}
            Data compared with the best fitting models $\Lambda$CDM 
            (solid) and  OCDM (dotted), see table \ref{table}. } 
\end{center}
\end{figure}

%%%%%%%%%%%%%%%%%%
We have fitted our model (PS and $T-M$ and $L_x-T$) to 
the following data sets. The cluster mass function of 
Bahcall \& Cen (1993), the temperature function of 
Henry \& Arnaud (1991), the luminosity function of Ebeling et al. (1997), 
and the flux function of Rosati et al. (1998) and De Grandi et al. (1999), 
see fig. \ref{fig_best_4curves}.   
The fit of the model to these data sets was performed using a Bayesian 
estimator discussed in Lahav et al. (1999) and as it was shown in Diego et al. 
(2000) with this estimator the bias in the estimation of the cosmological 
parameters is very small. A marginalization of the probability over the 
cosmological parameters $\Omega$ and $\sigma_8$ has shown that by 
combining the different data sets it is possible to drastically reduce 
the degeneracy between these two parameters, see fig. \ref{fig_Sigma8_Omega}.
The best model (see table \ref{table}) was found to be a low 
density universe with $\Omega \approx 0.3$ and $\sigma_8 \approx 0.8$.
The first three parameters in table \ref{table} are the cosmological 
ones and they appear in the PS formula for the mass function. 
The other parameters are for the $T-M$ and $L_x-T$ relations.
\begin{equation}
  T_{gas} = T_0 M_{15}^{\alpha}(1 + z)^{\psi} ,
  \label{Tx}
\end{equation}
where $M_{15}$ is the cluster mass in $h^{-1} 10^{15} M_{\odot}$ and, 
\begin{equation}
  L_x^{Bol} = L_0 M_{15}^{\beta}(1 + z)^{\phi} .
  \label{Lx}
\end{equation}
The constraints found for these relations are consistent with 
experimental determinations of these scalings. However we found 
some discrepancy in the $\alpha$ parameter mostly due to a bias 
in our estimator. See, however, Diego et al. (2000) for a detailed 
discussion on these parameters.\\

\noindent  
Both the flat $\Lambda$CDM and the OCDM ($\Lambda = 0$) models were 
found to be in good agreement with the data (see fig. \ref{fig_best_4curves}).
In fact both models listed in table \ref{table} are undistinguishable 
and they reproduce the data with almost the same good fit.

\noindent
The fact that both models are undistinguishable is because in our fit 
we have used low redshift data (see Diego et al. 2000 for a brief description 
of each data set). To distinguish the $\Lambda$CDM and OCDM models it is 
necessary to explore the cluster distribution as a function of redshift.  
This will be done with more or less success by undergoing X-ray experiments 
(CHANDRA, Newton-XMM) but the real improvement in the redshift (and sky) 
coverage of galaxy cluster surveys will come with the near future mm surveys. 
Through the SZE (Sunyaev-Zeldovich effect, Sunyaev \& Zeldovich 1972) 
it will be possible to 
go deeper in redshift than with X-rays due to the different behaviour 
of the selection functions in both bands, see Diego et al. (2001).
In that work, we have  shown how future measurements of the SZE can be used 
to constrain the cosmological parameters. 
\begin{table}
\large
\caption{Best $\Lambda$CDM  and OCDM models.}
\label{table}
\normalsize
%%%%%%%%\begin{flushleft}
\begin{center}
\begin{tabular}{lcccc}
\hline
%\hline
Parameter  & OCDM & $\Lambda$CDM \\
\hline
%\hline
$\sigma_8$     		      & $0.8^{+0.1}_{-0.1}$ & $0.8^{+0.2}_{-0.1}$\\
\hline
$\Gamma$      		      & $0.2^{+0.05}_{-0.1}$ & $0.2^{+0.05}_{-0.1}$\\
\hline
$\Omega_m$     		      & $0.3^{+0.2}_{-0.1}$ & $0.3^{+0.2}_{-0.1}$\\
\hline
$T_0 (10^8 h^{\alpha} K)$     & $1.1^{+0.1}_{-0.2}$ & $1.1^{+0.2}_{-0.3}$\\
\hline
$\alpha$       		      & $0.8^{+0*}_{-0.15}$ & $0.75^{+0.05*}_{-0.1}$\\
\hline
$\psi$        		      & $1.0             $ & $1.0$\\
\hline
$L_0 (10^{45}h^{\beta}h^{-2}erg/s)$  & $0.9^{+0.6}_{-0.0}$ & $1.5^{+0.3}_{-0.9}$\\
\hline
$\gamma$  		      & $3.1^{+0.2*}_{-0.3}$ & $3.2^{+0.1*}_{-0.4}$\\
\hline
$\phi$			      & $3.0^{+0}_{-2}$ & $1.0^{+2}_{-0}$\\
\hline
\end{tabular}
%%%%%%%%%\end{flushleft}
\label{table}
\end{center}
\end{table}
In particular we have concentrated 
on the Planck Surveyor mission which will explore all the sky at different 
mm frequencies and will detect above 30000 clusters with fluxes above 
30 mJy (see Diego et al. 2001 for an explanation of this limiting flux). 
This huge number 
of clusters would provide a valuable information about the cluster 
population. The constraints on the cosmological parameters will increase 
significantly by fitting the model to Planck observations. However, 
this information alone will not be enough to distinguish some models like 
those listed in table \ref{table}. \\

\begin{figure}
\begin{center} 
\psfig{figure=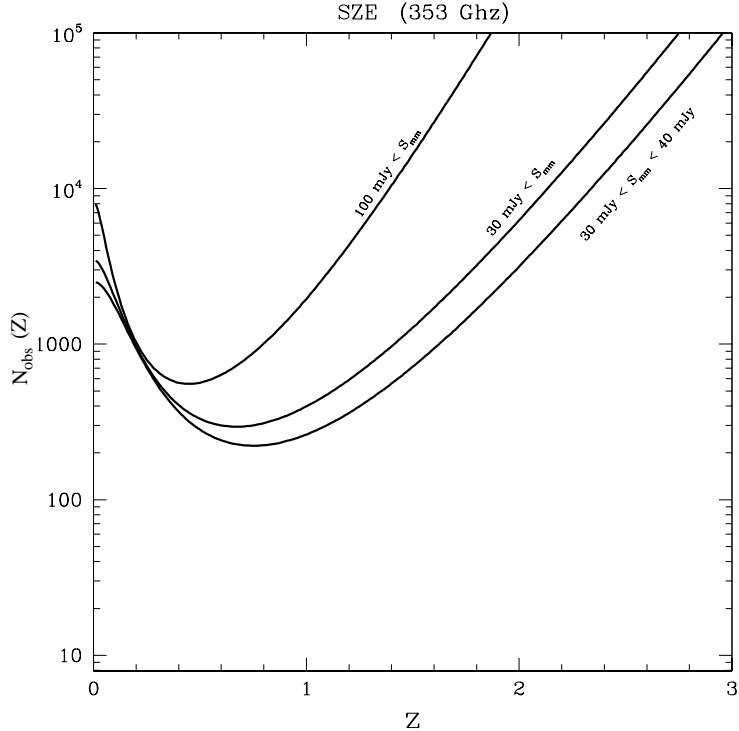,height=4.0in} 
\caption{\label{N_obs}
	  Number of clusters to be observed to distinguish  $N(>z)$ 
	  for the OCDM and $\Lambda$CDM models at a $3\sigma$ level.} 
\end{center}
\end{figure}

\begin{figure}
\begin{center} 
\psfig{figure=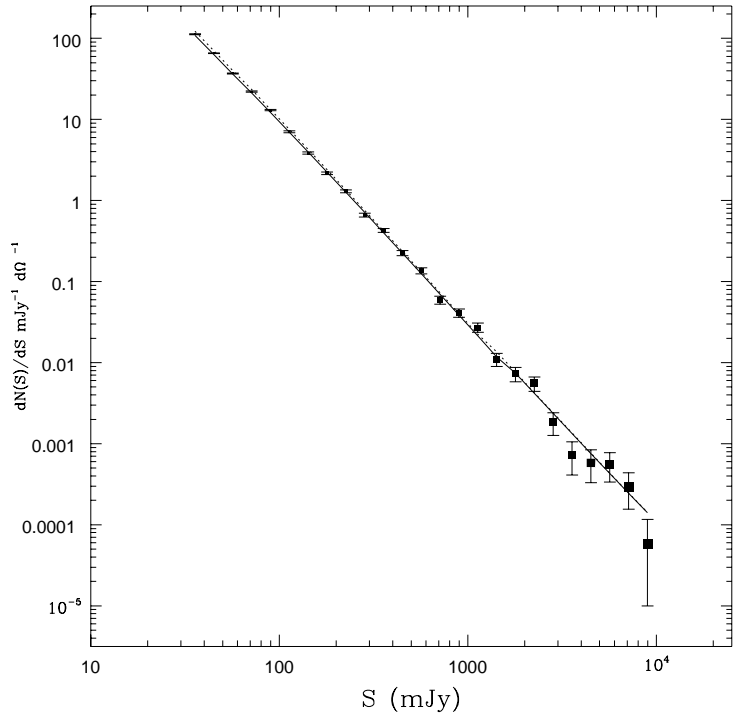,height=4.0in} 
\caption{\label{NS_Planck}
          $dN/dS$ curve for the $\Lambda$CDM (solid) and 
          OCDM (dotted) models in table \ref{table}} 
\end{center}
\end{figure}

\begin{figure}
\begin{center} 
\psfig{figure=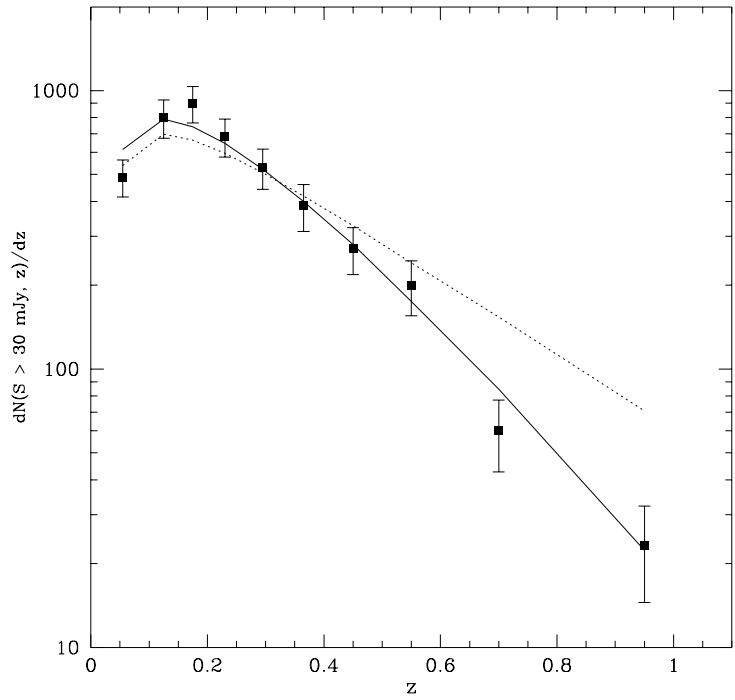,height=4.0in} 
\caption{\label{Nz_Planck}
	    $dN(S > 30  mJy, z)/dz$ (353 GHz) curve for the $\Lambda$CDM 
	    (solid) and OCDM (dotted) models in table \ref{table}. }

\end{center}
\end{figure}

\noindent
Despite the capabilities of Planck to observe galaxy clusters 
trough the SZE, it will be unable to determine their redshifts and therefore 
many models will predict the same cluster population as a function of the 
SZ flux. In fig. \ref{NS_Planck} we can see that both models would remain 
undistinguishable if only the information about the fluxes of the clusters 
is provided.\\
To distinguish those models it is required some knowledge about 
the cluster population as a function of redshift.  It will be needed, 
therefore, an optical follow up of some clusters (if their redshifts 
have not yet been determined in other wavebands). 
The number of clusters expected in Planck data  
is too large to make the follow up of all of them and consequently the 
cosmological studies based on the evolution of the cluster population 
must be restricted to a small subsample of the whole catalogue.
As an application, we have calculated the minimum number of clusters 
for which we should determine their redshift in order to distinguish 
the previously 
{\it undistinguishable} models listed in table \ref{table}. 
In that calculation we have 
required that the number of clusters above a given $z$ must differ 
at least $3\sigma$ for both models. The result is shown in 
figure \ref{N_obs}  for three different selection criteria of the clusters. 
In each one of the 3 lines we show the total number of clusters 
randomly selected from the  catalogue (with the only condition that 
the total flux must be $S_{mm} > 100 $ mJy top, $S_{mm} > 30 $ mJy middle and 
$30 $ mJy $< S_{mm} < 40 $ mJy bottom)  
which should be optically observed 
in order to distinguish (at a 3$\sigma$ level) $N^O$ and $N^{\Lambda}$  
at redshift $z$ (i.e. the total number of observed clusters needed to 
have a $3 \sigma$ difference in $N(>z)$ for the two models). 
As can be seen from the figure, by determining the redshift of about 300 
clusters randomly selected from the whole Planck catalogue (fluxes 
above 30 mJy) it would be possible to distinguish 
the models listed in table \ref{table} by looking at the different 
behaviour of the cluster population about redshift $z \approx 0.6$. 
To illustrate how it is really possible to distinguish them, 
in fig. \ref{Nz_Planck} we show the behaviour of both models and the 
comparison with simulated data (corresponding to the $\Lambda$CDM model 
and for a survey covering a sample of about 300 clusters). 
It is evident that by looking at the evolution of the cluster population 
as a function of redshift (for a small subsample of clusters) it would 
be possible to distinguish both models.\\

\noindent
The previous calculation tell us that an analysis based on a small subsample 
of the whole Planck catalogue (with $z$, fig. \ref{Nz_Planck}) 
could distinguish models which are undistinguishable with recent X-ray 
data and future Planck data (no $z$, fig. \ref{NS_Planck}).\\ 

\begin{figure}
\begin{center} 
\psfig{figure=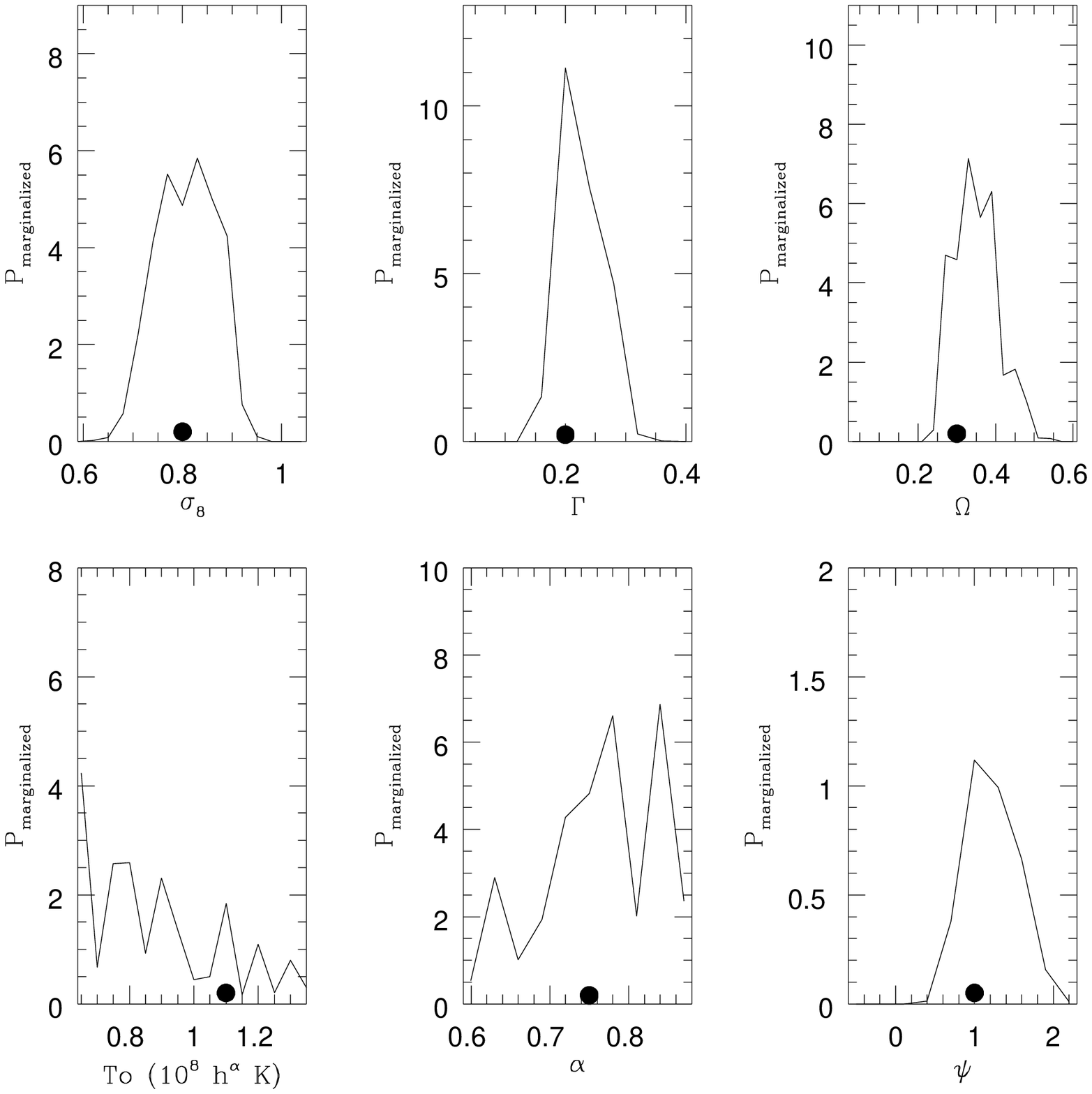,height=4.0in} 
\caption{\label{fig_Margin}
	  Marginalized probability in each parameter. 
          Black dots represent the input model 
          ($\Lambda$CDM model in table \ref{table}).} 
\end{center}
\end{figure}

\noindent
By looking at figures \ref{NS_Planck} and 
\ref{Nz_Planck} we see that a powerful analysis should make full use of both 
data sets since some models would be excluded by the first data set and 
others would be excluded by the second one. Following 
Diego et al. (2000) we combined both data sets in order to see how well 
could be constrained the cosmological parameters with these future data 
sets. Also following the same work we have considered that the $T-M$ 
relation was a free parameter one. The result of this combination 
of data sets is shown in fig. \ref{fig_Margin}. As can be seen the recovery 
of the cosmological parameters is very good. A joint fit to the Planck 
$N(S)$ curve and an optically identified $N(z)$ curve for a small subsample 
of clusters of the Planck catalogue will allow an independent test of the 
cosmological parameters. This result is almost independent of the assumed 
amplitude $T_0$ and scaling exponent $\alpha$ in the 
$T-M$ relation as it was discussed in Diego et al. (2001). However the 
method is sensitive to the choice of the redshift exponent $\psi$.\\

\noindent
We can also conclude that the inclusion in the model of the cluster 
scaling relations ($T-M$ and $L_x-T$) as free parameter ones it is 
important for the modeling of low redshift data since many models 
can fit the low redshift data and the assumption of one or another 
scaling in the $T-M$ and/or $L_x-T$ can favor some models to the detriment 
of others.\\
On the contrary, when fitting data which includes the cluster evolution 
with redshift, the evolution of the cluster population with redshift is 
dominated by the cosmological model and the cluster scaling relations play a 
secondary role. However, the redshift dependence of these relations (as can 
be appreciated in fig. \ref{fig_Margin}, $\psi$ parameter) should be 
carefully considered in this case.

%%%%%%%%%%%%%%%%%%%%%%%%%%%%
\section*{Acknowledgments} %
%%%%%%%%%%%%%%%%%%%%%%%%%%%%

This work has been supported by the 
Spanish DGESIC Project  
PB98-0531-C02-01, FEDER Project 1FD97-1769-C04-01, the 
EU project INTAS-OPEN-97-1192, and the RTN of the EU project   
HPRN-CT-2000-00124. \\
JMD acknowledges support from a Spanish MEC fellowship FP96-20194004 
and financial support from the University of Cantabria.

%%%%%%%%%%%%%%%%%%%%%%%
\section*{References} %
%%%%%%%%%%%%%%%%%%%%%%%

\end{document}